# Hybrid colour filters for multispectral imaging

Xin He[1], Yajing Liu[1], Paul Beckett[2], MD Hemayet Uddin[3], Ampalavanapillai Nirmalathas[1] and Ranjith Rajasekharan Unnithan[1]

1 Department of Electrical and Electronic Engineering, The University of Melbourne, Melbourne, VIC, 3010, Australia. Correspondence and requests for materials should be addressed to R.R.U.
(*email: r.ranjith@unimelb.edu.au)
2 School of Engineering, RMIT University, Melbourne, VIC, 3000, Australia.
3 Melbourne Centre for Nanofabrication, Australian National Fabrication Facility, Clayton, VIC, 3168, Australia.

Abstract:

Multispectral cameras capture images in multiple wavelengths in narrow spectral bands. They offer advanced sensing well beyond normal cameras and many single sensor based multispectral cameras have been commercialized aimed at a broad range of applications, such as agroforestry research, medical analysis and so on. However, the existing single sensor based multispectral cameras require accurate alignment to overlay each filter on image sensor pixels, which makes their fabrication very complex, especially when the number of bands is large. This paper demonstrates a new filter technology using a hybrid combination of single plasmonic layer and dielectric layers by computational simulations. A filter mosaic of various bands with narrow spectral width can be achieved with single run manufacturing processes (i.e., exposure, development, deposition and other minor steps), regardless of the number of bands.

# 1. Introduction

Multispectral camera (MC) technology has the ability to collect information well beyond human eyes or normal colour cameras. This is because human eyes and conventional colour cameras receive an image in three colour bands (red, green and blue (RGB)) with broad spectral widths. Hence, it is difficult to distinguish fine spectral features within the image captured across the bands due to their overlap. For example, the spectral width in the conventional cameras is around 80 nm to 100 nm for each of the RGB bands. Multispectral cameras based on a single image sensor capture images in multiple wavelengths using narrow band filters fitted on each image sensor pixel [1-4]. It is possible to extract additional spectral information otherwise not possible from these multiple images [1-11]. The spectral width of each band is measured at FWHM of the spectral distribution, where FWHM is defined as the wavelength range at which the received optical power is half of its maximum.

MC technology sub-divides the information collected by the image sensor into narrow bands defined by the small FWHM values of its constituent filters. The filtered information can then be grouped, processed and highlighted to show a large contrast with its reference background. Examples are already abound across diverse applications. In gastroenterology research, each observed case (i.e., tissue type) may relate to a particular range of electromagnetic radiation or to a different reflectance spectrum. In this case, ulceration of the colon has been shown to exhibit a larger reflectance at the boundary between red and NIR wavelengths (around 600nm to 1000nm) [5] and can therefore be made to appear much brighter against a non-diseased background. In the pathology area, the location of veins can be difficult to detect for many reasons, such as dark skin, excess loss of blood or even extreme dehydration. To solve this problem, MC been applied as either a combination of a commercial camera and a NIR bandpass filter or a standard camera together with a NIR illuminator. As NIR light will easily penetrate the skin, medical staff are able to more easily to identify the exact position of veins [6]. In agroforestry research, it is known that diseased or stressed plants reflect more NIR light than healthy leaves. Thus, MC systems have been installed on drone aircraft to fly over the crops, observing harvest conditions in real time [7].

There are three different types of MC system currently on the market. The first uses multiple cameras with a single narrow-band filter in front of each camera [8-9]. This type of system can be easily customized simply by replacing the spectral filter with one appropriate to the specific application. The main disadvantage is that these multiple camera platforms tend to be bulk and power hungry, which limits their use [10]. Moreover, each camera has their own imaging viewpoint which creates co-registration problems [11-12]. While second MC systems with single camera plus a rotating optical filter disk can eliminate the viewpoint disparity problem, they are still quite bulky and introduce time shifts that complicate the processing of moving images, such as when the system is integrated with a drone for agroforestry applications [9].

The third type of MC available on the market is based on a single camera sensor called single sensor based multispectral camera (SSMCs) [1-4]. While the filter technologies employed in these systems are almost universally commercial-in-confidence, the limited information available in the open literature (e.g., [4]) indicates that the filter wavelength is typically tuned by changing the filter material and/or its thickness. This kind of technology is also described in many recent research papers [14-15] and patents [16-19]. It is worth mentioning that the photodetector under each pixel collects the optical signal after it has passed through both a micro-lens and the filter layer(s) [20]. There are typically two ways to install a filter array on top of an image sensor. It can be either fabricated onto a substrate and then aligned with the image sensor, or the array can be directly fabricated onto the surface of the sensor itself. Either method requires a large number of separate nanofabrication fabrication processes [22-24] for each individual spectral band, such as pixel-to-pixel mask alignment, UV exposure, and filter installation. Therefore, the fabrication can become extremely complicated for making a filter mosaic with required number of bands, particularly in the case of hyperspectral filters where the number of filters (bands) may reach into the hundreds [4, 10]. The filter technology makes the SSMCs very expensive and hence limits their usage in wide applications.

This has triggered research on new filter technologies suitable for SSMCs. The two main requirements of the filter technology are developing technologies with less complex fabrication process and getting spectral width (FWHM) narrow for bands. There has been much activity aimed at improving the FWHM of the bands, as evidenced by recent patents and published papers [14-15, 16-21, 25-33]. By far the smallest FWHM reported to date in the visible/NIR range is approximately 0.11nm observed only over 10 nm wavelength range in a dielectric grating structure [27]. A metallic grating structure has also been demonstrated in which an aluminium grating built on an $Al_2O_3$ buffer layer, in turn located on a quartz substrate, results in a minimum FWHM of 20nm [29]. Metal-Dielectric-Metal (MDM) or even MDMDM metallic gratings have also been designed and optimized to exhibit good FWHM values of under 30nm [30]. However, as these metallic grating structures have been shown to be polarization dependent, SSMCs based on them may fail to collect an optical signal if the incident light is in an opposite polarization direction.

Metallic hole array based multispectral filters have been proposed [10, 25-27] that show FWHM value smaller than 50nm. Although these multispectral filters have been applied to colour and multispectral imaging applications [10], they are unable to operate cover the whole visible and NIR range. Si nanowire based narrow bandpass filter arrays have also been integrated on a monochrome image sensor for multispectral imaging, but with mechanical scanning and have a FWHM greater than 50nm [28].

In this paper, we demonstrate a new filter technology suitable for multispectral imaging using a hybrid combination of single plasmonic layer and dielectric multilayers. The hybrid filters can produce spectral width (FWHM) values smaller than 50nm, with a minimum FWHM as low as 17nm at some wavelengths. Moreover, these proposed multispectral filters are easily tuneable and a multiband filter mosaic can be manufactured using a single stage maskless fabrication process regardless of the number of spectral bands, including one-time exposure, chemical development and filter deposition. We have demonstrated two filter structures as examples, one exploiting localized surface plasmons resonances and other using surface plasmon resonances using the proposed hybrid topology to demonstrate their operation in the visible and NIR wavelengths by computational simulations.

**2. Design of Hybrid Multispectral Filters**
**2.1. Angle Independent Narrow Bandpass Filters based on the Localized Surface Plasmon Resonances**

In the first of the proposed filters presented here, Aluminium (Al) disks structures were investigated. Previously, these have been used to develop subtractive colour filtering as the different resonant wavelengths (valley wavelengths) depend largely on the disk diameter [36]. Subtractive primary colours typically include cyan, magenta and yellow, and are the inverse transmission spectra of red, green and blue respectively. This work has resulted in the multispectral filter mosaic (RGB colour filter plus NIR filter) using Al disk in a square array using the proposed hybrid technique; Al disc (plasmonic layer using localized surface plasmons) and dielectric layers (silicon nitride ($Si_3N_4$) and spin-on-glass (SOG)). An initial narrow-band filter was designed and optimized using COMSOL software [37, 38].

We found by appropriately arranging the Al disks in a square array as shown in **Fig. 1(a),** the peak transmission wavelength can be tuned by varying the diameter of the Al disk, while the optimal spacing between two adjacent disks was found to be around 1/10th of the peak wavelength. In these simulations, the substrate is quartz with a refractive index of 1.5. A 250nm thick $Si_3N_4$ layer with a refractive index of 2 was deposited as a buffer layer on the quartz. Onto this buffer layer, a thin plasmonic layer of 50nm Al disks with refractive index [39] was designed to tune the peak wavelength. Lastly, a coating of 200nm Spin on glass (SOG) covered the sample. The general 3D structure of the filter is shown in Fig. 1(b). PMLs were applied at the top and bottom, SBCs layers to the remaining sides of the simulation model. Port boundary conditions with propagation constants of $2\pi \times 1.42/\lambda$ for SOG and $2\pi \times 1.5/\lambda$ for quartz were used between the top PML and SOG as well as between the bottom PML and the quartz. Periodic boundary conditions (PBC) were applied on the four sides of the middle Al disk (unit cell as shown in Fig 1 (a) dotted lines) and $Si_3N_4$ block.

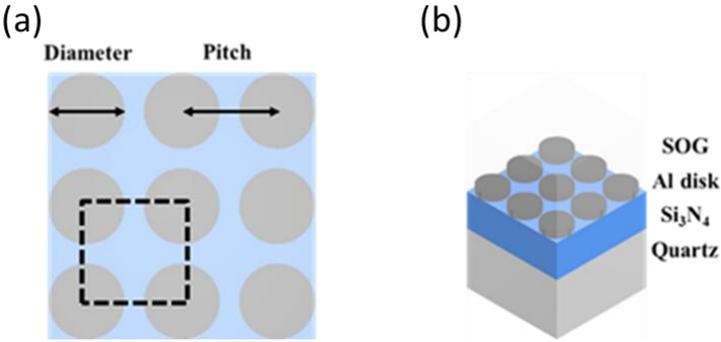

Fig. 1. Schematic of the proposed narrow-band filter based on the hybrid Al disk (a) top view of the filter (dotted line shows the unit cell used in the simulation) (b) 3D representation of the filter structure with the plasmonic layer (Al disc) and dielectric layers ($Si_3N_4$ and SOG).

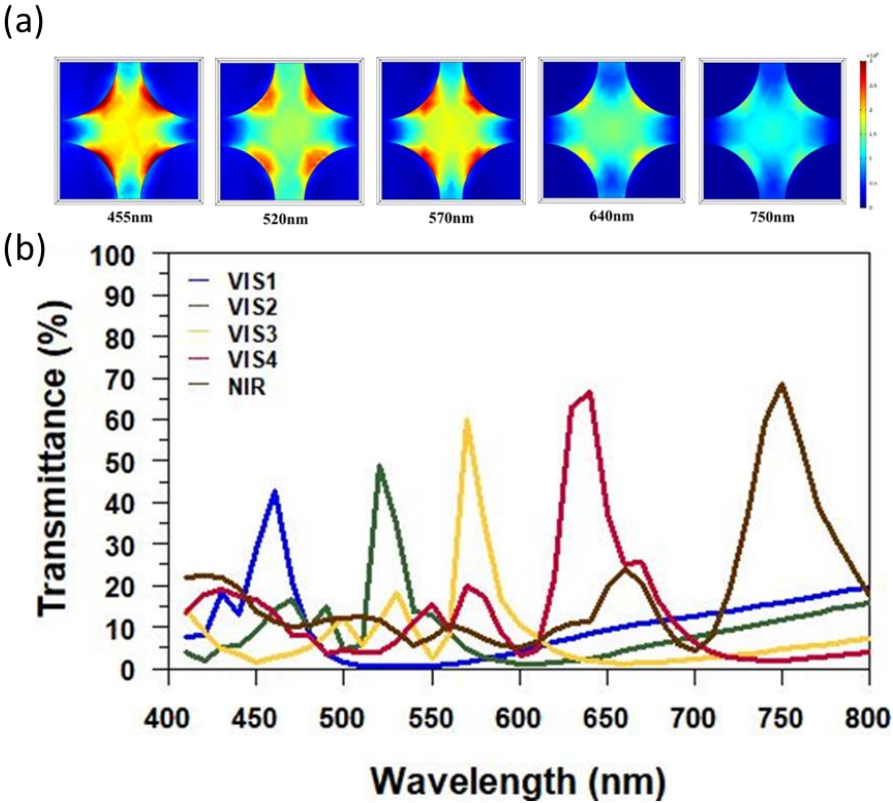

Fig. 2. (a) Normalized electric field of localised surface plasmons for blue (460 nm), green (520 nm), yellow (570 nm), red (635 nm) and NIR (750 nm) (b) Transmission spectra of hybrid Al disk based multispectral filters with respect to the wavelengths mentioned in (a).

The transmission efficiency was calculated using s-parameter, $|S_{21}|^2$. Fig 2 (a) shows the normalized electric field at peak wavelengths for blue (460 nm), green (520 nm), yellow (570 nm), red (635 nm) and NIR (750 nm) showing the localized surface plasmons. The resulting FWHM for green and yellow (Fig. 2b) are about 18nm and 15nm, respectively, with transmission efficiency values larger than 45%. The wavelength tuning was mainly achieved by tuning diameter of the disk (Figure. 1(a)). Table 1 shows the optimized filter parameters along with their optical characteristics.

Table 1. Results of blue, green, yellow, red and NIR narrow bandpass filters.

| Parameter | Blue | Green | Yellow | Red | NIR |
|---|---|---|---|---|---|
| Peak Wavelength (nm) | 460 | 520 | 570 | 635 | 750 |
| Period (nm) | 240 | 280 | 310 | 350 | 420 |
| Disk Diameter (nm) | 200 | 220 | 260 | 280 | 340 |
| Peak Transmission | 42% | 48% | 60% | 65% | 68% |
| FWHM (nm) | 25 | 18 | 15 | 30 | 45 |

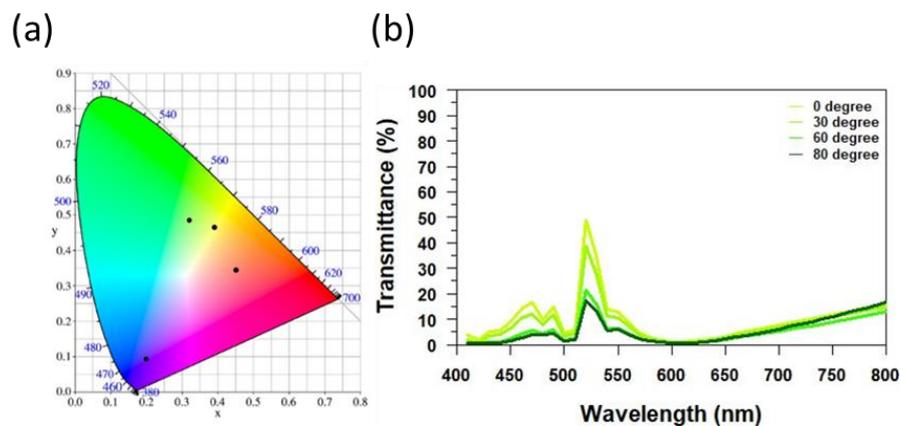

Fig. 3. (a) CIE chromaticity chart for green, blue, yellow and red (b) Transmission vs Wavelength over FFOV: 0° ≤ Angle of Incidence ≤ 80°

Fig 3(a) shows the CIE chromaticity chart overlaid with the transmission data of blue, green and red showing that the filter values are falling in the appropriate part of the color space. The transmission efficiency with respect to angle of incidence was investigated using the blue filter (blue, 460nm) as shown in Fig. 3. The full field of view (FFOV) values varied from 0° (i.e., normal to the surface) to 80° off-axis. It was observed that the peak wavelength position not shifted with respect to the angle with minimum reduction in the transmission intensity up to 30o. Furthermore, when the FFOV is 80°, the blue filter may adversely affect the optical response of the NIR filter above 800nm.

It can be seen from Table 1 that the proposed hybrid filter geometry made of Al nano-disks and a $Si_3N_4$ layer exhibits FWHM values smaller than 50nm in the visible and NIR wavelength and is therefore a promising candidate for making multispectral filter mosaic. Moreover, the proposed narrow-band filter behavior is independent of the angle of incidence and, as all the structures (substrate, $Si_3N_4$, Al, SOG) have the same base thickness, the array requires only a single-exposure fabrication process without any mask alignment.

## 2.2. Metallic Nanohole Array Integrated on a Dielectric Multilayer for IR Multispectral Imaging

The second of the two structures presented in this paper is a hybrid filter illustrated in Fig. 4 using surface plasmon resonances. This filter is made up of a plasmonic filter on a dielectric multilayer on top of a quartz substrate (refractive index 1.5) covered with Spin on Glass with refractive index 1.42 (Fig. 4). A dielectric multilayer is formed from a layer of $SiO_2$ with refractive index ($n_1$) of 1.45 surrounded by two $TiO_2$ layers with larger refractive indexes ($n_2$) of 2.1. The thickness of the $TiO_2$ layers are both 190nm, which is a quarter of the target NIR wavelength of 760nm. The $SiO_2$ layer is determined by the equation $d_1n_1=d_2n_2$, where $d_1$ and $d_2$ are the thicknesses of the $SiO_2$ and $TiO_2$ layers, respectively, and $n_{1/2}$ are the corresponding refractive indices, resulting in a $SiO_2$ thickness of approximately 280nm. This part of the filter on its own behaves as an anti-reflection layer with a passband from approximately 650nm to 1000nm as illustrated by the transmission spectrum (Fig. 5.) of the dielectric multilayer before the plasmonic layer is added. The plasmonic filter comprises an array of nano-holes in a hexagonal arrangement [40] inserted in the 150nm Au film. Fig. 5 (a) shows the normalized electric field of the surface plasmons in the hole array (top view) at peak wavelengths for NIR 1-5.

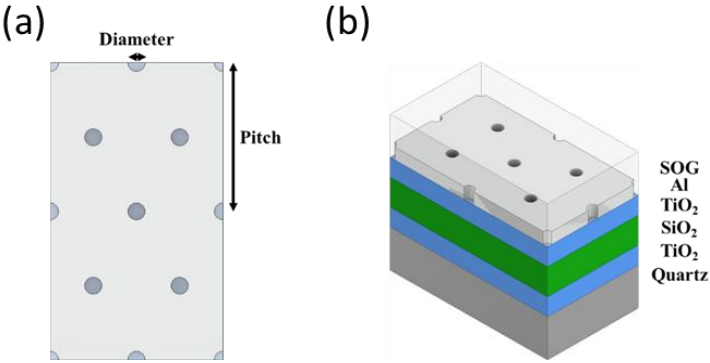

Fig. 4. Nanostructure of the proposed hybrid IR multispectral filter (a) top view (b) 3D structure of the filter structure (Al – Plasmonic layer and $TiO_2$, $SiO_2$, SOG are dielectric layers)

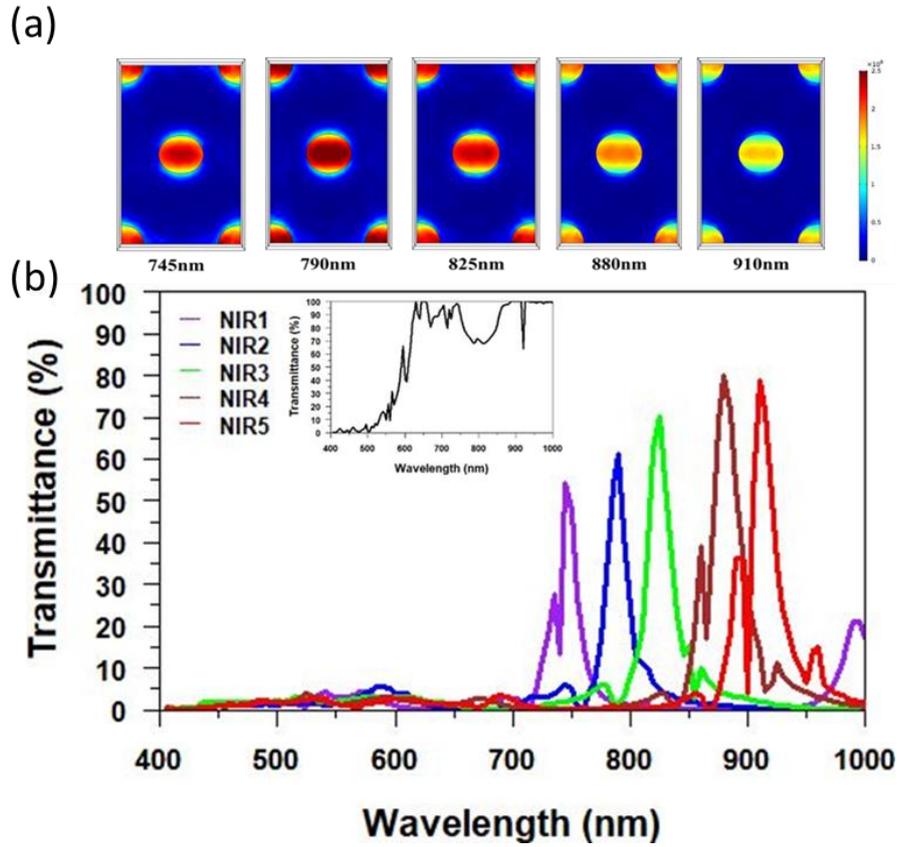

Fig. 5. (a) Normalized electric field of the surface plasmons in the hole array at peak wavelengths for NIR 1-5 (b) The simulated transmission spectrum of the hybrid IR multispectral filters. The inset shows the transmission spectrum of $TiO_2$-$SiO_2$-$TiO_2$ dielectric multi-layer without the plasmonic layer.

Table 2. Results of Au hole array integrated with $TiO_2$-$SiO_2$-$TiO_2$ multilayer.

| Peak wavelength (nm) | 745 | 790 | 825 | 880 | 910 |
|---|---|---|---|---|---|
| Period (nm) | 480 | 520 | 550 | 600 | 630 |
| Hole Diameter (nm) | 180 | 200 | 220 | 240 | 240 |
| Peak transmission efficiency (%) | 55 | 60 | 70 | 80 | 78 |
| FWHM (nm) | 20 | 20 | 22 | 25 | 20 |

The transmission efficiency and the resulting spectrum of the hybrid filter is plotted in Fig. 5 (b). The inset of Fig. 5 (b) shows the spectrum from $TiO_2$-$SiO_2$-$TiO_2$ dielectric multilayers without the plasmonic layer. The FWHM and peak transmission efficiency results for the device are presented in Table 2. As shown in Table 2, the FWHM of a multispectral filter in the NIR wavelength can be less than 25nm (typically around 20nm) with a best-case transmission efficiency of 80%. The dielectric multilayer remains the same in each case, showing that the peak wavelength can be tuned by simply adjusting the period between adjacent holes. However, hole array based filters are limited in the angle independant performance due to the SPPs. Colour filters based on LSPs are more angle independent than SPPs based ones. We have carried out further simulations to study the effect of non-infinite filter size on the transmission coefficient and spectral width by replacing the periodic boundary conditions with semi-infinite boundary condition. The transmission efficiency and spectral width are almost the same for the metallic disk based structure due the LSP, where one single disc can produce colour. But hole array requires interference of surface plasmons and hence multiple holes are required for producing a colour. For the hole array based structure, the spectral width and transmission efficiency were reduced as the pixel size decreased. This requires covering the hole array based filter on at least 2x2 pixels on the image sensor with pixel size of 3.3 um along with suitable optics to produce required spectral width and acceptable transmission.

In conclusion, this paper has presented a multispectral filter technology using a hybrid combination of single plasmonic layer and dielectric layers by computational simulations. The hybrid filter technology has reduced the fabrication complexity to a single stage process for making any number of bands in a filter mosaic with narrow spectral widths and easy wavelength tuneability. We have demonstrated working of the proposed hybrid design in the visible and NIR using two filter examples, one using localized surface plasmons resonances and other surface plasmon resonances. This proposed technology will significantly reduce the fabrication complexity and cost of multispectral filter mosaic compared to the existing complex SSMC filter techniques. The current technology will further make the SSMCs applied in wider applications in precision agriculture, medicine, forestry, night vision and remote sensing.

**Acknowledgements**
The authors acknowledge financial support from the Australia Research Council under Discovery Project: DP170100363.

**Conflict of interests**
The authors declare that they have no conflict of interest.